\documentclass[12pt,preprint]{aastex}

\def\beq{\begin{equation}}
\def\eeq{\end{equation}}

\def\cm{\,{\rm {cm}}}

\def\Mpc{\,{\rm {Mpc}}}
\def\kpc{\,{\rm {kpc}}}

\def\kms{\,{\rm {km\, s^{-1}}}}
\def\msun{\,{\rm M_\odot}}
\def\vcir{V_{\rm c}}
\def\v200{V_{200}}

\def\E51{\,{\rm E_{51}}}

\def\S13{S_{-13}}

\def\pc{\,{\rm pc}}

\usepackage{amsmath,color,epsf}

\def\kms{\,{\rm {km\, s^{-1}}}}

\begin{document}

\title{Metallicity and HI Column Density Properties of
Damped Lyman-$\alpha$ Systems }

\author{J.L. Hou$^1$, C.G. Shu$^{2,3,1}$, S.Y. Shen$^1$, R.X. Chang$^1$,
W. P. Chen$^2$, \& C.Q. Fu$^{1}$}

\affil{$^1$Shanghai Astronomical Observatory, CAS, 80 Nandan Road,
Shanghai 200030, CHINA \\
$^2$Graduate Institute of Astronomy, National Central University,
Taiwan, CHINA\\
$^3$Joint Center for Astrophysics, Shanghai Normal University,
Shanghai, 200234, CHINA}

\email{hjlyx@shao.ac.cn}

\begin{abstract}

Based on the disk galaxy formation theory within the framework of
standard $\Lambda$CDM hierarchical picture (Mo, Mao \& White
1998), we selected modelled damped Lyman-alpha systems (DLAs),
according to their observational criterion $N_{\rm HI} \ga
10^{20.3} \cm^{-2}$ by Monte Carlo simulation with the random
inclinations being considered, to examine their observed
properties. By best-fitting the predicted metallicity distribution
to the observed ones, we get the effective yield for DLAs about
$0.25Z_{\odot}$, which is comparable to those for SMC and LMC. And
the predicted distribution is the same as that of observation at
the significant level higher than 60\%. The predicted column
density distribution of modelled DLAs is compared with the
observed ones with the corresponding number density, gas content
being discussed. We found that the predicted number density $n(z)$
at redshift 3  agree well with the observed value, but the gas
content $\Omega_{DLA}$ is about 3 times larger than observed since
our model predicts more DLA systems with higher column density. It
should be noted that the predicted star formation rate density
contributed by DLAs is consistent with the most recent
observations if the star formation timescale in DLAs is assumed to
be 1 $\sim$ 3 Gyr. Meanwhile, the connection between DLAs and
Lyman Break galaxies is discussed by comparing their UV luminosity
functions which shows that the DLAs host galaxies are much fainter
than LBGs. We also predict that only few percent of DLAs can host
LBGs which is also consistent with current observations. However,
there is a discrepancy between model prediction and observation in
the correlation between metallicity and HI column density for
DLAs. We suggest that this could result from either the inadequacy
of Schmidt-type star formation law at high redshift, the
diversities of DLA populations, or the model limitations. Although
our current simple model cannot fully reproduce the observed DLA
velocity distribution, we argue that such kind of model can still
provide valuable information for the natures of DLAs.

\end{abstract}

\keywords{Galaxies: formation and evolution $-$ Quasars:
absorptions $-$ Galaxies: DLAs}

\section{INTRODUCTION}

The nature of clouds (or protogalaxies) that host high redshift
Damped Lyman-$\alpha$ systems (DLAs) has been a controversial
topic for many years. By definition, DLAs are absorbers seen in
quasar optical spectra with HI column densities $N_{\rm HI} \ge
10^{20.3} \cm^{-2}$. The studies on DLAs include the evolution of
neutral gas (Storrie-Lombardi \& Wolfe 2000; P$\acute{e}$roux et
al. 2002), their metal abundance and enrichment history (Pettini
et al. 1994, 1997a, b; Lu et al. 1996; Ma \& Shu 2001; Prochaska
\& Wolfe 2002; Prochaska 2003), the dust depletion (Vladilo 1998,
2002; Hou, Boissier \& Prantzos 2001) and their kinematics
(Prochaska \& Wolfe 1997, 1998; Wolfe \& Prochaska 2000). They
have constituted powerful tools to investigate the properties of
distant galaxies (or of their building blocks), and consequently
provided strong clues to the formation and evolution of galaxies.

It has been a common knowledge that DLAs are the progenitors of
present-day galaxies. But substantial debate continues over
exactly what populations of galaxies are responsible for them.
Based on observed internal kinematics of DLAs traced by the
associated heavy element absorption lines, Prochaska \& Wolfe
(1997, 1998) proposed that DLAs relate to rotating-support
galactic disks at the time before which substantial gas
consumption has taken place. Meanwhile, direct imaging shows that
low and intermediate redshift DLA host galaxies span a variety of
morphological types from dwarf, irregular, and low surface
brightness (LSB) to normal spiral galaxies (Le Brun et al. 1997;
Rao \& Turnshek 1998; Kulkarni et al. 2000, 2001, Chen \& Lanzetta
2003; Rao et al. 2003). Numerical simulations and semi-analytic
models have shown that such observational signatures can be
explained by a mixture of rotation, random motions, infall, and
mergers of proto-galactic clumps (Haehnelt, Steinmetz \& Rauch
1998; Maller et al. 2001, 2003). Moreover, several authors have
suggested that high redshift DLAs could be associated with Lyman
Break Galaxies (LBGs)(Nulsen, Barcos \& Fabian 1998; Shu 2000;
Schaye 2001b) since strong galactic winds from LBGs would also
give rise to DLAs observed against background QSOs. Although
recent work done by Moller et al. (2002) did not suggest that
there be a DLA/wind connection, further observational evidence is
still needed.

In theory, one of the methods for DLAs research is to assume that
DLAs are galaxies with different morphological types, such as
disks, irregulars, dwarfs, or LSBs, then to compare model
predictions with the observed properties, such as abundance
patterns, kinematics or number densities (Matteucci, Molaro \&
Vladilo 1997; Meusinger \& Thon 1999; Prantzos \& Boissier 2000;
Hou, Boissier \& Prantzos 2001; Ma \& Shu 2001; Mathlin et al.
2001; Calura, Matteucci \& Vladilo 2003; Boissier,
P$\acute{e}$roux \& Pettini 2003; Lanfranchi \& Friaca 2003).
Another approach is to start from the framework of cosmic
structure formation and evolution. Hence the observed DLA
properties are strong tests for various cosmological models and
also for galaxy formation and evolution models (Mo \&
Miralda-Escude 1994; Gardner et al. 1997, 2001; Cen et al. 2003;
Nagamine, Springel \& Hernquist 2003; Cora et al. 2003; Okoshi et
al. 2004). It should be pointed out that semi-analytic models
(SAMs) have been quite successful in understanding galaxy
formation and evolution (Baugh et al. 1998; Mo, Mao \& White 1998;
Somerville \& Primack 1999; Somerville, Primack \& Faber 2001).
This technique adopts statistical methods to follow the growth of
dark matter halos, which is a major concept of the building-up
structure in the universe. By further introducing some physical
rules, one could fairly describe the gaseous and stellar process
within dark halos (Kauffmann 1996; Mathlin et al. 2001).

However, when this kind of models was applied to DLAs, it is found
that model results cannot simultaneously reproduce the observed
kinematical data and column density data (Haehnelt et al. 1998;
Jedamzik \& Prochaska 1998; Maller et al. 2001). Some alternative
explanations have been proposed such as multiple gas discs along
the line of sight or a large contribution of mergers at high
redshift (Churchill et al. 2003).

We notice that in a recent paper done by Boissier, P\'eroux \&
Pettini (2003), some DLAs properties, such as number densities,
column density distribution and gas densities, were discussed
based on a simple model. Many interesting results were obtained.
For instance, they claimed that in order to reproduce the observed
properties, LSBs and spirals are at least responsible for DLAs,
while dwarf galaxies may not be dominant (but see Efstathiou
2000).

In the present paper, we will adopt a SAM to examine in detail the
observed metallicity, HI column density and star formation
properties of DLAs in the context of standard hierarchical picture
(White \& Rees 1978; White \& Frenck 1991) assuming that DLAs are
hosted by disk galaxies. The disk galaxy formation model with
single disks is adopted because we mainly concentrate on HI column
densities and the cosmic star formation rate density contributed
by DLAs rather than their kinematics (see more detailed
discussions in Section 3).

In Section 2, we describe our galaxy formation model in the
$\Lambda$CDM cosmogony with the considerations of how star
formation and chemical enrichment proceed. The simulated DLA
sample is selected by Monte Carlo simulation according to their
observational criterion. We compare the model predictions with
observed DLA properties in Section 3 on the following items:
metallicity, column density, number density, neutral gas content,
contributed star formation rate density, and the correlation
between metallicity and HI column density. The discussions of the
model parameters on the results are also presented in this
section. Main conclusions are summarized in Section 4. As an
illustration, DLA properties are assumed at redshift $z\sim 3$ and
the following $\Lambda$CDM cosmogony is adopted throughout the
paper with $\Omega_0 = 0.3$, $\Omega_{\Lambda} = 0.7$, $H_0 = 100h
\kms \Mpc^{-1}$ and $\sigma_8 = 0.9$. Whenever a numerical value
of $h$ is needed, we take $h = 0.7$.

\section{MODELS}
\subsection{Galaxy formation}

The galaxy formation model in the present paper comes from that
for disk galaxies suggested by Mo, Mao \& White (1998, hereafter
MMW), in which the primordial density fluctuations give rise to
galactic halos, and baryons within individual halos condenses
later and forms disks due to their angular momentum. The relation
between halo mass $M$ and its circular velocity $\vcir$ is given
by
\begin{equation} \label{eq:halomass}
M = \frac {\vcir^3}{10 G H(z)},
\end{equation}
where  $G$ is the gravitational constant, $H(z)$ is Hubble
constant at redshift $z$. Disks are assumed to be thin, to be
centrifugal balance, and to have exponential surface profiles
\begin{equation} \label{eq:exdisk}
 \Sigma (R) = \Sigma_0 \exp (-R/R_d),
\end{equation}
where $\Sigma_0$ and $R_d$ are, respectively, the central surface
density and the scale length, and they can be expressed as
\begin{eqnarray} \label{eq:Sigma0}
\Sigma_0 \approx 384h^{-1}\msun\pc^{-2}
\left(\frac{m_d}{0.05}\right)
                \left(\frac{\lambda}{0.05}\right)^{-2}\nonumber\\
                \left(\frac{\vcir}{250\kms}\right)
                \left [\frac{H(z)}{H_0} \right]
\end{eqnarray}
 and
\begin{equation} \label{eq:Rd}
 R_d \approx 8.8h^{-1}\kpc \left(\frac{\lambda}{0.05}\right)
\left(\frac{ \vcir}{250\kms} \right) \left[
\frac{H(z)}{H_0}\right]^{-1},
\end{equation}
respectively. Here $m_d$ is the mass ratio of disk to halo,
$\lambda$ is the halo dimensionless spin parameter, $H_0$ is
Hubble constant at present day, respectively (see MMW for
details). The disk global properties are uniquely determined by
$\vcir$, $\lambda$, $m_d$ and $H(z)$, while other cosmological
parameters, such as $z, \Omega_0$, and $\Omega_{\Lambda}$, affect
disks only indirectly through $H(z)$. Since Hubble constant $H(z)$
increases with redshift, it is expected from the above equations
that galaxy disks of any given $\vcir$ and $\lambda$ are less
massive and smaller but more compact at higher redshift.

The density profile of a halo is assumed to be the NFW profile
(Navarro, Frenk \& White 1997), with the concentration $c=10$
which is the median value for its distribution in CDM cosmogony
(Jing 2000). The rotation speed at radius $r$ corresponding to
this profile is
\begin{equation} \label{eq:VH}
V_H^2(r) = \vcir^2 \frac{1}{x}\frac{{\rm ln}(1+cx)-cx/(1+cx)}
                 {{\rm ln}(1+c)-c/(1+c)},
\end{equation}
where $x = r/r_{200}$, and $r_{200} = \vcir / 10H(z)$ is the
virial radius of a halo.

The halo mass function at any redshift $z$ is described by the
Press-Schechter formalism (Press \& Schechter 1974):
 \begin{eqnarray} \label{eq:PS}
  {\rm d}N = -\sqrt{2 \over \pi}{\rho_{0} \over
M}{\delta_{c}(z) \over \Delta(\overline R)} {{\rm
d}\ln\Delta(\overline R) \over {\rm
d}\ln M} \nonumber \\
{\exp}\left[-{\delta_{c}^{2}(z) \over {2\Delta^{2}(\overline
R)}}\right]{{\rm d} M \over M},
 \end{eqnarray}
where $\delta_{c}(z)=\delta_{c}(0)(1+z)g(0)/g(z)$ with $g(z)$
being the linear growth factor at $z$ and $\delta_{c}(0)\approx
1.686$, $\Delta(\overline R)$ is the linear $rms$ mass fluctuation
in top-hat windows of radius $\overline R$ which is related to the
halo mass $M$ by $M=(4\pi/3){\overline \rho}_{0}\overline R^{3}$,
with ${\overline\rho}_{0}$ being the mean mass density of the
universe at $z=0$. A detailed description of the PS formalism and
the related cosmogonic issues can be found in the Appendix of MMW.

The distribution function of spin parameter is always assumed to
be independent of time and $\vcir$. It can be well described by a
lognormal function
\begin{equation} \label{eq:spin}
p(\lambda) {\rm d}\lambda = \frac{1}{\sqrt{2\pi}\sigma_\lambda}
                \exp \left[-\frac{{\rm ln}^2(\lambda/\bar{\lambda})}
                {2\sigma_\lambda^2}\right ] \frac{{\rm
                d}\lambda}{\lambda},
\end{equation}
with $\bar{\lambda} = 0.05$ and $\sigma_\lambda = 0.5$ (Warren et
al 1992; Lemson \& Kauffmann 1999), respectively.

Although the baryon fraction within individual galactic halos is
usually treated as a constant initially, the effective fraction of
baryonic mass which can form disks should be different from
galaxies to galaxies due to supernovae feedback (Somerville \&
Primack 1999; Baugh et al. 1999; Cole et al. 2000). Since small
galactic halos have shallow potential wells which will lead to
strong mass loses due to galactic winds and mass outflows (Shu, Mo
\& Mao 2003), $m_d$ should be a function of $\vcir$ which can be
expressed as (Dekel \& Silk 1986; White \& Frenk 1991)
\begin{equation} \label{eq:md}
m_d = \frac{m_{d0}}{1+(\frac{\vcir}{150 \kms})^{-2}},
\end{equation}
where $m_{d0}$ is the maximum baryon fraction within halos. Here
we take $m_{d0}$ = 0.1 according to the cosmic nucleosynthesis
(Burles \& Tytler 1998) and the corresponding discussions of this
parameter on the model results are in Sec. 3.7.

It should be pointed out that the interaction between disks and
bulges is not considered in the present paper. This effect will be
important for very compact objects which correspond to galaxies
with $\lambda \leq 0.025$. Because the fraction of these galaxies
is less than 10\% of the whole galaxy population
(eq.{\ref{eq:spin}), and DLAs are dominated by extended galaxies
with large $\lambda$ due to large absorbtion cross sections, this
treatment is reasonable and will not influence our results (see
Sec. 3.2).

\subsection{Star formation and chemical evolution}

With the cumulation of gas in the disk, star formation takes
place. According to Kennicutt (1998) with the consideration of
rotation velocities in disks, the adopted star formation
prescription in the present paper is
\begin{eqnarray} \label{eq:SFR}
\Psi= 0.1 \left(\frac{\epsilon_0}{0.1}\right)
\left[\frac{\Sigma_g(R)}{\msun\pc^{-2}}\right]^{1.5}
\left[\frac{V_{rot}(R)}{220\kms}\right]\nonumber \\
\left(\frac{R}{8.5\kpc}\right)^{-1}  \msun {\rm Gyr^{-1}}\pc^{-2},
\end{eqnarray}
where $\Psi$ is the star formation rate (SFR) per unit area,
$\Sigma_g(R)$ is the gas surface density and $V_{rot}(R)$ is the
rotation speed at disk radius $R$, $\epsilon_0$ is the star
formation efficiency which is set to be 0.1 based on the disk
modelling of Boissier \& Prantzos (2000). The SFR prescription in
the present paper is somewhat different from that in Ma \& Shu
(2001), within which the disk instability was taken into account.
Note that eq. (\ref{eq:SFR}) has been very successful in modelling
the Milky Way disk properties both locally and globally, as well
as other galaxies  both at local universe and at high redshifts
(Boissier \& Prantzos 1999, 2000; Prantzos \& Boissier 2000).
Especially, such prescription is necessary for models to reproduce
the observed abundance gradient in the disks of the Milky Way and
external galaxies (Hou, Prantzos \& Boissier 2000; Henry \&
Worthey 1999). For the Milky Way type galaxies, eq. (\ref{eq:SFR})
is equivalent to $\Psi \propto \Sigma_g^{1.5}R^{-1}$ since
$V_{rot}(R)$ keeps nearly constant within very wide ranges in
disks and outskirts.

The rotation speed contributed by an exponential disk $V_D$ is
(Freeman 1970)
\begin{equation} \label{eq:VD}
V_D^2(R) = V_{d0}^2 y \left [I_0(y)K_0(y)-I_1(y)K_1(y) \right ],
\end{equation}
with $y = R/2R_d$ and $V_{d0} = (2\pi G \Sigma_0 R_d)^{1/2}$.
$I_n$ and $K_n$ are modified Bessel functions of order $n$,
respectively. Then the resulted rotational velocity $V_{rot}$ for
an exponential disk within a NFW halo can be calculated through
$V_{rot}^2 = V_H^2+V_D^2$. Whenever the disk mass fraction of a
galaxy with the given $\vcir$ is known, its rotation velocities at
different radius $R$ in the disk is determined. The star formation
prescription eq. (\ref{eq:SFR}) can be applied at different
radius.

As mentioned above, we focus on the DLA properties of HI column
densities, metallicities and star formation rates but their
kinematics, we will not apply detailed prescriptions of either gas
infall or outflow. Instead, the effects of gas infall and outflow
are included within the obtained effective yield (see below) and
the disk mass fraction $m_d$ (eq. {\ref{eq:md}}), respectively. No
radial gas inflow or outflow within disks is considered.

Under the approximation of instantaneous recycling, the chemical
evolution in disks can be expressed by the simple closed-box model
(Pagel 1997) with metallicity $Z$ to be
\begin{equation} \label{eq:Z}
Z-Z_i = -y \, {\rm ln}\mu,
\end{equation}
where $Z_i$ is the initial metallicity of gas and is assumed to be
$0.01Z_\odot$, $y$ is the effective yield, and $\mu$ is the gas
fraction. The evolution of gas surface density $\Sigma_g$ is
determined by
\begin{equation} \label{eq:Sigmat}
\frac{{\rm d}\Sigma_g(R,t)}{{\rm d}t} = -(1-f_R)\Psi(R,t),
\end{equation}
where $f_R$ is the return fraction of stellar mass into ISM, and
we take  $f_R = 0.3$ for a Salpeter stellar initial mass function
(see Madau, Pozzetti \& Dickinson 1998).

According to their detailed analysis, Bechtold et al. (1998) and
Lanfranchi \& Friaca (2003) pointed out that star formation
proceeds within DLAs in a typical timescale $1 \sim 3 \rm Gyr$
(see also Dessauges-Zavadsky et al. 2004). In the present paper,
the star formation timescales for individual DLA galaxies are
reasonably chosen to be random between 1 to 3Gyr, and initially
($t = 0, z=3$) the gas surface density distribution of a galactic
disk with a given $\vcir$ is $\Sigma_{g0}(R) = \Sigma_0 \exp
(-R/R_d)$ (eq. \ref{eq:exdisk}). It should be pointed out that the
effect yield in our model is obtained by best-fitting the modelled
metallicity distribution of DLAs to that of observations. Shorter
star formation timescale adopted will lead to higher effect yield
and vice versa. We will come back to the discussions of the
effects for different star formation timescale intervals adopted
between 0.5 and 3Gyr on our model results in Sec. 3.7.

\subsection{Modelling DLA population}

Since the distributions of $\vcir$ and $\lambda$ for individual
halos are known, we can generate a sample of galaxies by a
Monte-Carlo simulation in the $\vcir$-$\lambda$ plane at $z\sim
3$. According to eqs. (\ref{eq:SFR}) and (\ref{eq:Z}), we can then
investigate star formation and chemical evolution for individual
disks based on their randomly selected star formation timescales.
The modelled DLAs are selected over the sampled galaxies by random
sightlines penetrating disks according to the observed selection
criterion, i.e., $N_{\rm HI} \gtrsim 10^{20.3} \cm^{-2}$. Here,
random inclinations for disks in the sky are considered and a
hydrogen fraction $x = 0.7$ is assumed with the consideration of
Helium in gas. We assume that the physical quantities, such as
column density, metallicity, SFR, of a modelled DLA, are
represented as the physical quantities at the point where the
sightline penetrates the disk, and the projected distance between
the point and the galactic center is named as the impact parameter
of the modelled DLA. We adopt $\vcir$ from 50 to $360\kms$, which
corresponds roughly to the observed range for spirals and
irregular galaxies at the present day. A lower limit of $\vcir$,
50 $\kms$, is chosen because gas in halos with $\vcir \lesssim
50\kms$ would not be cooled down to form disks due to the strong
external UV background at high redshift (Rees 1986).

\section{MODEL PREDICTIONS vs OBSERVATIONS}

\subsection{Observations}

DLAs have shown many observational properties, including
metallicities, column densities, kinematics, etc. The observed
metallicities of DLAs adopted in the present paper mainly come
from the recent compilation presented by Hou, Boissier \& Prantzos
(2001) and by Kulkarni \& Fall (2002) for the Zn element. All the
data presented by those authors are compiled from the results of
various observers. Moreover, one new observed DLA from P\'eroux et
al. (2002) and four from Prochaska et al. (2003b) are added. Since
our model focuses on DLAs at redshift $z \sim 3$, the observed
DLAs with redshift $z>2 $ are selected for the direct comparison
with our model results. We noticed that in the compilation of
Kulkarni \& Fall (2002), Zn abundances of one third DLAs are
represented as upper/lower limits. Here, we do not simply exclude
them. Instead we choose their limit values when we discuss the
observed metallicity distribution of DLAs. It should be pointed
out that the data without these limitations has the same
distribution function as that we do in 94\% significance level
after the K-S test.

Although including other non-refractory elements may be helpful in
enlarging the data sample, this may also introduce other
uncertainties, such as dust depletion, etc. Therefore, we prefer
to rely only on the Zn elements because Zn is generally believed
to be an undepleted or only mildly depleted element. But this
might cause completeness problems for the adopted metallicity
distribution due to incomplete Zn sample. To make this point
clearer, we have compared our Zn sample with more complete Fe
sample from Prochaska et al. (2003a) which is assumed to be in the
Galactic Halo depletion pattern (Savage \& Sembach 1996) since Fe
has a significant level of dust depletion. The K-S test shows that
the two samples have the same distribution function at
significance level higher than 97$\%$. This shows that our Zn
sample distribution is generally consistent with more complete
sample, and therefore reasonable to be compared with theoretical
model.

The observed data of HI column densities come from the survey of
Storrie-Lombardi \& Wolfe (2000) (hereafter SW00), within which 85
DLA absorbers have their column densities $N_{\rm HI} \gtrsim
10^{20.3}\cm^{-2}$ covering the redshift range from 0.008 to
4.694. Especially, about 73\% of DLAs have the redshift $z\ga 2$.
The observed number densities of DLAs come from P$\acute{e}$roux
et al. (2003). The observed contribution to the cosmic SFR density
by DLAs comes from the most recent work done by Wolfe, Gawiser \&
Prochaska (2003)(hereafter WGP03) according to the $\rm CII^*$
absorption lines.

\subsection{$\vcir$, $\lambda$ and impact parameter}

\begin{figure} [t]
\resizebox{\hsize}{!}{\includegraphics{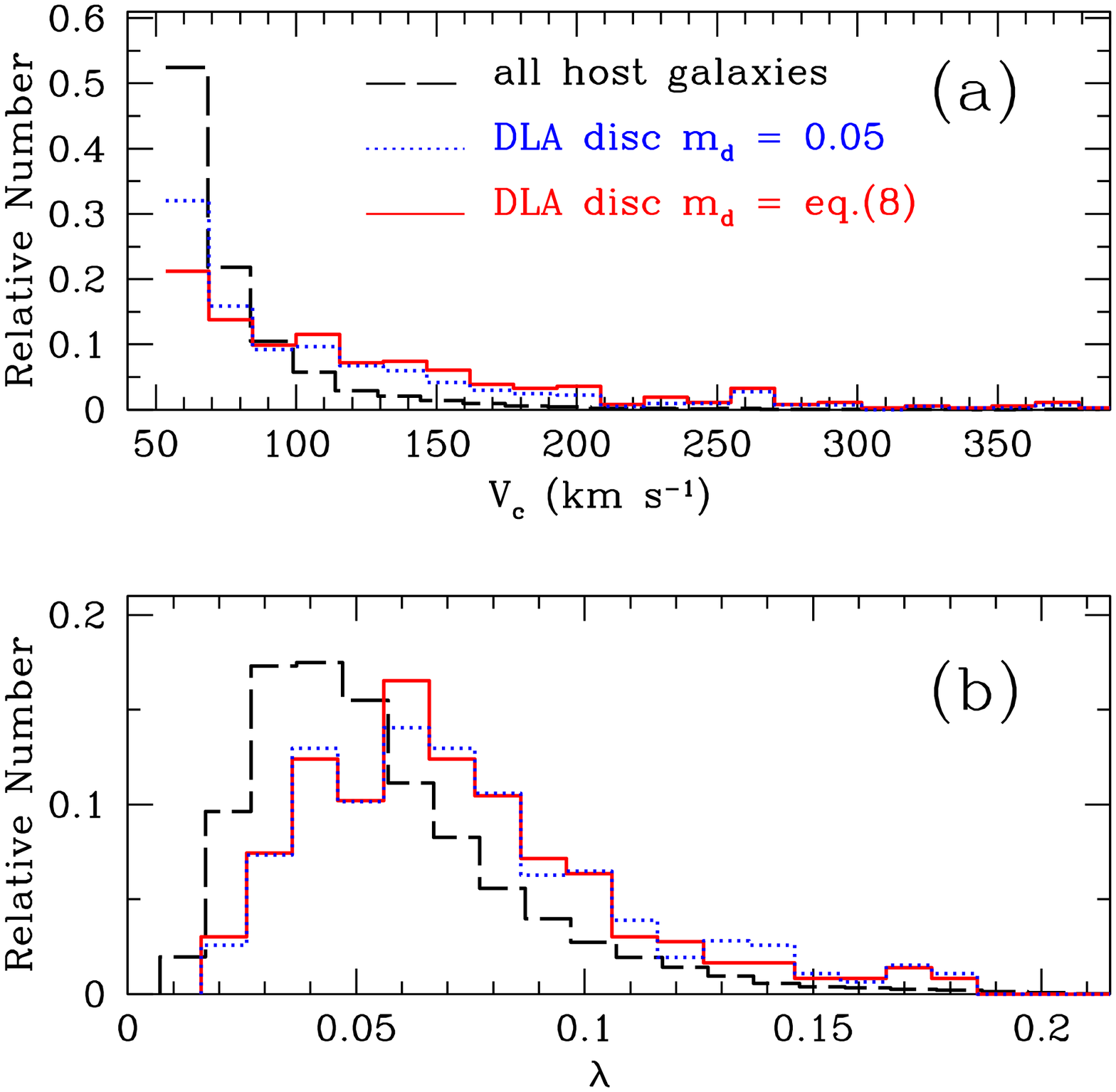}}
\figcaption[fig1.eps]{The distributions of $\vcir$ and $\lambda$
with the solid, long-dashed and dotted histograms denoting the
results of the modelled DLAs, all galaxies, and the results with
constant $m_d = 0.05$, respectively.} \label{Fig1}
\end{figure}

\begin{figure} [t]
\resizebox{\hsize}{!}{\includegraphics{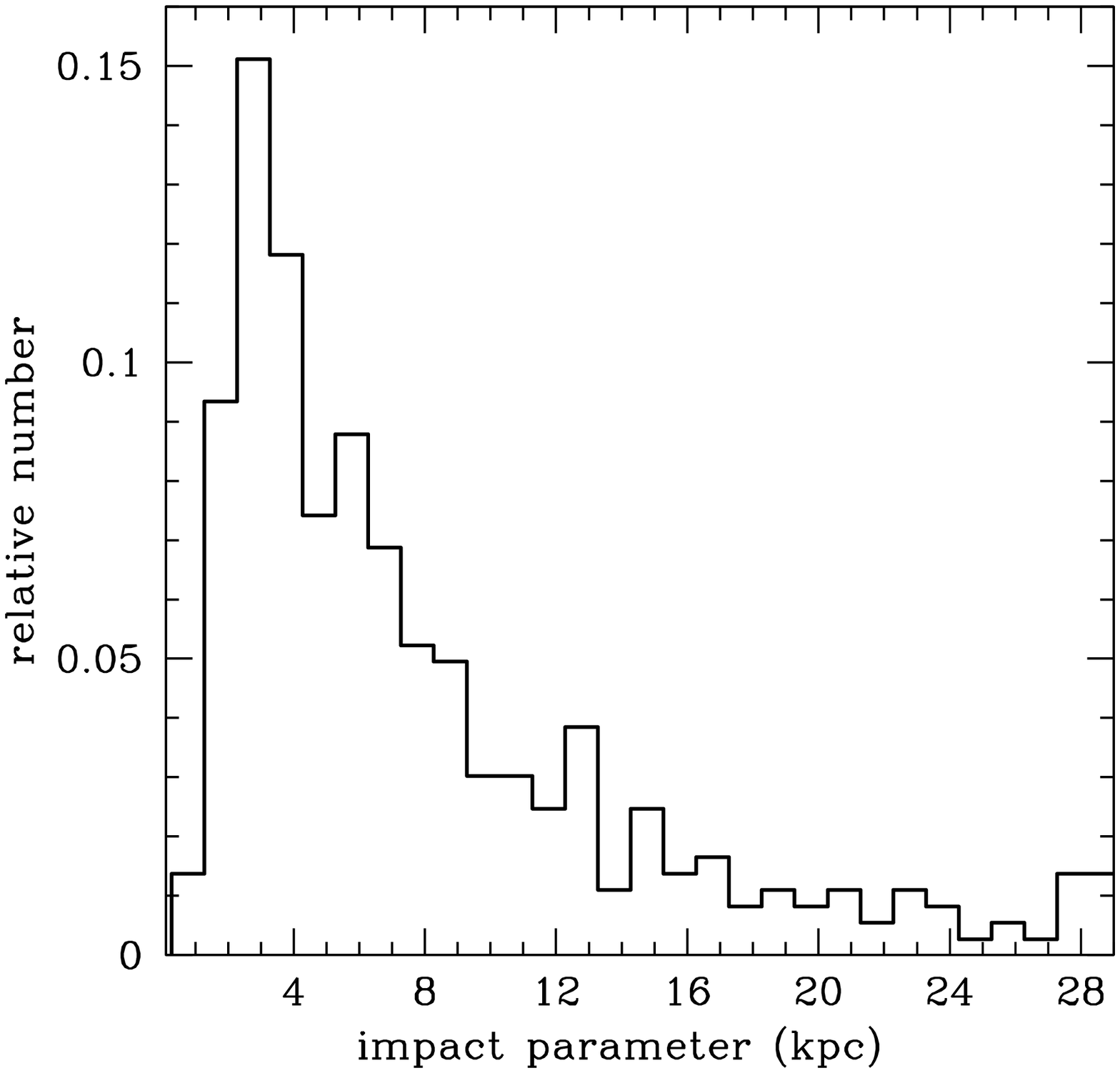}} \figcaption {The
distributions of impact parameters of modelled DLA populations.}
\label{Fig2}
\end{figure}

The predicted distributions of $(\vcir, \lambda)$ for the modelled
DLA population, which are selected according to the criterion
described in Sec. 2.3, are plotted in Fig. 1 as solid histograms.
As for comparison, we also plot the corresponding distributions of
all galaxies predicted by eqs. (\ref{eq:PS}) and (\ref{eq:spin})
in long-dashed histograms and the results with fixed $m_d = 0.05$
in dotted histograms (see Sec. 3.7 below), respectively. It can be
found as expected that modelled DLAs are still dominated by small
galaxies because the number of small galaxies is very large
although individual absorption cross-sections are small.

The predicted halo mass function for DLAs is much flatter than
that for all galaxies predicted by PS formalism, because larger
galaxies, although less numerous, are easier to be selected as
modelled DLA hosts due to their larger absorption cross-sections.

The $\lambda$ distribution of selected DLAs peaks around 0.065
with the median value of 0.08, larger than those for all galaxies
predicted by eq. (\ref{eq:spin}). This means that DLA hosts in our
model are bias to extended disk galaxies, in consistence with MMW.
This also implies that our ignorance of interaction between disk
and bulge is acceptable since the effect is important only for
galaxies with small $\lambda$ ($<0.025$).

Moreover, the resulted distribution of the impact parameters of
selected DLAs is shown in Fig. 2. It is found that the peak is
around 3kpc, which is resulted from the huge amount of small halos
in PS formalism and the finite radius for individual galaxies that
can produce DLAs estimated by eq. (14) in MMW. Because the DLAs
are dominated by small galaxies which are always faint, the peak
implies that the host galaxies of DLAs are difficult to be
observed photometrically.}

It should be pointed out that the $\vcir$ distribution in our
model seems to be not fully consistent with the observed kinematic
characteristics of DLAs, which is very common in current SAMs
(Prochaska \& Wolfe, 1997). Maller et al. (2001) proposed some
alternatives, such as gas discs could be more extended or DLAs
sightlines could pass through multiple gas discs, i.e.,
substructures, in a parental halo. Indeed, those improved models
could fairly reproduce the observed kinematics.

Because we mainly focus on column density, metallicity and SFR of
DLA population as emphasized above, our current single disc model
could still provide valuable insights into the DLA properties.

\subsection{Metallicity Distribution and Effective Yield}

\begin{figure} [t]
\resizebox{\hsize}{!} {\includegraphics{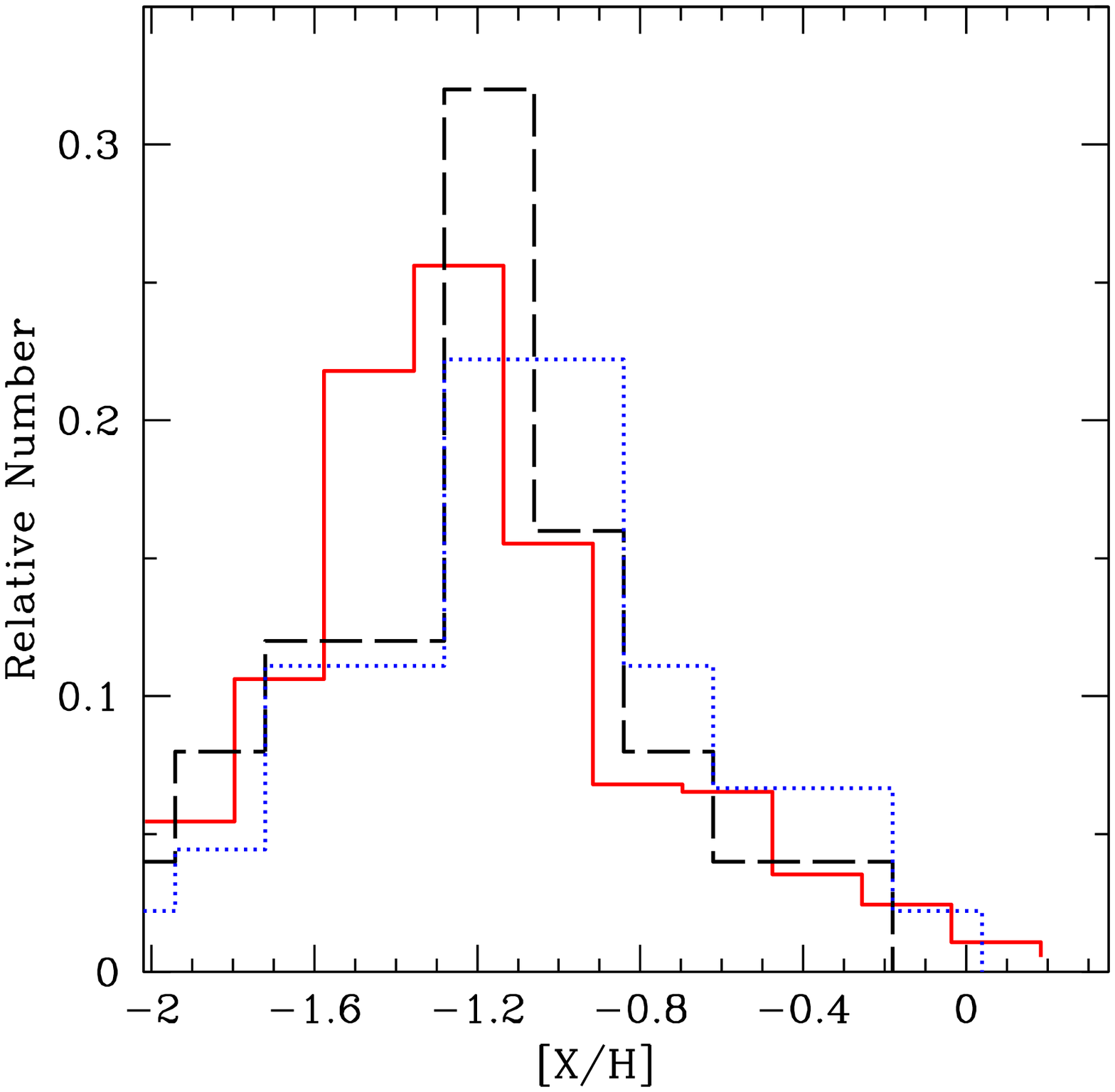}} \caption{The
metallicity distributions with the solid, dashed and dotted
histograms denoting the model prediction, observed DLAs with $z>2$
and all DLAs, respectively (see text for details).}
\label{Fig3}
\end{figure}

As we know, dust depletion will underestimate the true
metallicities of DLAs if only based on their observed spectrum.
Unfortunately, the physical prescription for the dust depletion is
still poorly known although there are some models available in
this field. To avoid the complicated dust depletion problems, we
adopt the Zn abundance as the metallicity indicators for DLAs
because Zn is usually regarded as the undepleted element in the
interstellar medium (ISM) (Savage \& Sembach 1996) although its
exact nucleosynthesis nature is still unclear. Moreover, the Zn
abundance is a good tracer of iron from the disk to the halo in
the Milky Way galaxy (e.g. [Zn/Fe] is close to zero everywhere).
However, recent abundance measurements in the very metal poor
stars have found some Zn enhancements indicating a possible Type
II supernova origin (Prochaska et al. 2000; Umeda \& Nomoto 2002;
Nissen et al. 2004).

The predicted metallicity distributions of DLAs is plotted as a
solid histogram in Fig. 3 while the observed distributions of DLAs
with $z>2$ and all DLAs are plotted as dashed and dotted
histograms, respectively. It should be noted that the predicted
distribution is obtained by best-fitting the model result to the
observational distribution for DLAs with $z>2$ through tuning the
effective yield $y$ in eq. (\ref{eq:Z}). We get the effect yield
$y = 0.25Z_\odot$ for the best-fit result with the assumption of
star formation timescale for DLAs being random between 1 and 3Gyr,
which is consistent with both the observational basis and
theoretical prescriptions (Bechtold et al. 1998; Lanfranchi \&
Friaca 2003; Dessauges-Zavadsky et al. 2004). Although the $N_{\rm
HI}$-weighted mean metallicity is a cosmological measure, we do
not discuss this quantity here because of large difference between
theoretical predictions and observations as Prochaske et al
(2003a) pointed out. Indeed, our model shows that the $N_{\rm
HI}$-weighted mean metallicity is about $-0.79$, higher than the
observed values, which is about $-1.21$ for our Zn sample (limits
not included). This could be resulted by the discrepancy between
theoretical results and observations for the correlation between
[Zn/H] and $N_{\rm HI}$, which will be discussed in Sec. 3.6.

The Kolmogorov-Smirnov method is applied to best-fit the modelled
distribution to the observed ones with $z>2$, which shows that the
two data sets follow the same population distribution function at
the significance level higher than 60$\%$. We conclude that the
predicted metallicity distribution agrees well with the high
redshift observations. Moreover, from the figure we can find that
the metallicity distribution of observed DLAs with $z>2$ is very
similar to that of all observed DLAs (Pettini et al. 1997a;b). But
the mean metallicity for $z>2$ DLAs seems to be a bit poorer,
which implies that the mean metallicity of DLAs might evolves with
cosmic time (Kulkarni \& Fall 2002, Prochaska et al. 2003a).

The obtained effective yield $y = 0.25 Z_{\odot}$ is similar to
those of SMC and LMC ($\sim$ 0.25$Z_\odot$, Binney \& Merrifield
1998) and of disk clusters ($\sim$ 0.30$Z_\odot$, Pagel 1987). The
low effective yield for DLAs, compared with that for the solar
neighborhood (about 0.7$Z_\odot$), implies that the star formation
precesses in DLAs should be less active than those in the Milky
Way-type galaxies. Because DLA hosts are dominated by small
galaxies for which gravitational potential wells are very shallow,
supernovae feedback should be significant (Efstathiou 2000; Shu,
Mo \& Mao 2003) which has been considered in eq. (\ref{eq:md})
while their star formation seems to be inactive.

\subsection{Column Density, Number Density and Gas Content}

The frequency distribution of HI column density $f(N_{\rm HI}, z)$
for DLAs , which is defined as the number of absorbers per unit
$N_{\rm HI}$ and per unit absorption distance $X$, is very
important for understanding galaxy formation and evolution in the
universe. It can be expressed as (see also Boissier, P\'eroux \&
Pettini 2003)
\begin{equation} \label{eq:fn}
f(N_{\rm HI},z) = \frac{n}{\Delta N_{\rm HI} \sum_i(\Delta X)_i},
\end{equation}
where $n$ is the number of DLAs with HI column density between
$N_{\rm HI}-\frac{1}{2}\Delta N_{\rm HI}$ and $N_{\rm
HI}+\frac{1}{2}\Delta N_{\rm HI}$ detected in the spectra of QSOs
encompassing a total absorption distance $\sum_{i} (\Delta X)_i $
from $z$ to $z + \Delta z$, and $dX = \frac{(1+z)^2}{E(z)}dz$ with
$E(z) = H(z)/H_0$. Based on large bright QSO surveys, Wolfe et al.
(1995), SW00, and recently P$\acute{e}$roux et al. (2003) have
derived the distribution function $f(N_{\rm HI},z)$. SW00 found
that $f(N_{\rm HI},z)$ can be well approximated by a power-law
function of HI column density, while Peroux et al. (2003) found
that the distribution of $f(N_{\rm HI},z)$ can be fitted with a
$\Gamma$ function down to Lyman Limit systems with the index of
$\beta \sim -1.0$.

\begin{figure} [t]
\resizebox{\hsize}{!}{\includegraphics{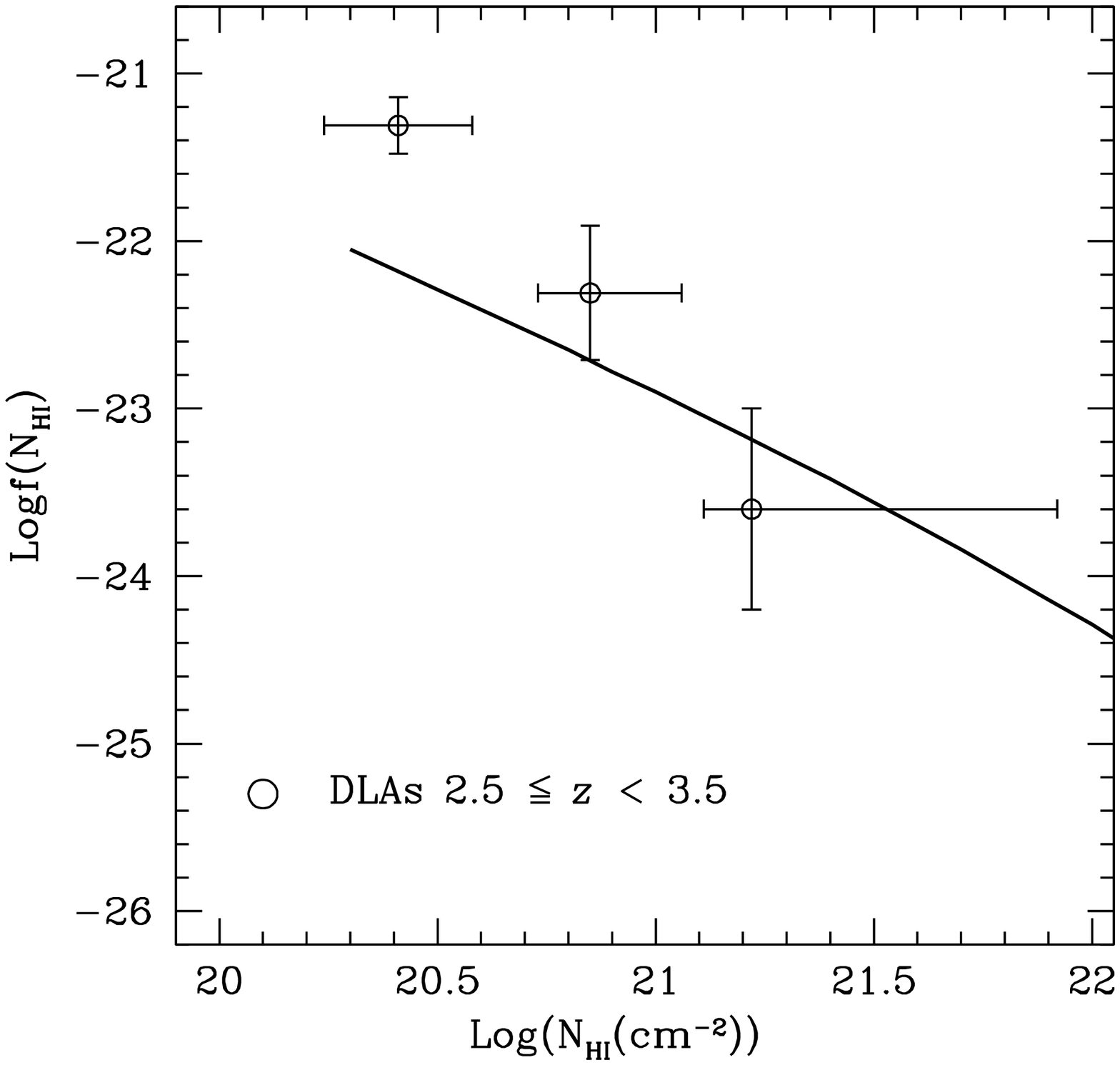}}
\caption{$f(N_{\rm HI})$ vs $N_{\rm HI}$ for DLAs with the solid
line denoting the model prediction at $z\sim 3$, and the circles
with error bars denoting the observations for $z = 2.5\sim 3.5$
which are taken from SW00.} \label{Fig4}
\end{figure}

We plot the model prediction of $f(N_{\rm HI},z)$ at $z \sim 3$ in
Fig. 4 as a solid line while the observed frequency distributions
for DLAs with $z = 2.5 \sim$ 3.5 from SW00 is plotted as circles
with error bars respectively. It can be found from the figure that
the predicted distribution agrees well with observations at the
high $N_{\rm HI}$ end. But it is smaller than observations at low
$N_{\rm HI}$ end with the maximum difference as large as
3$\sigma$, i.e., the predicted distribution is a bit flatter than
observed ones. The similar discrepancy appear in the more
complicated numerical simulation study of Nagamine, Springel \&
Hernquist (2003) as well. This could be due to the limitation of
our simple model or perhaps this could be a failing of the
$\Lambda CDM$ power spectrum which deserves further
investigations.

There exists an upper limit in both the observed and modelled
distributions of DLA HI column densities with the modelled one
being a bit larger. Observational bias has been proposed by a
number of authors in order to explain the difference
(Boiss$\acute{e}$ et al. 1998; Prantzos \& Boissier 2000; Schaye
2001a). As suggested by Boiss$\acute{e}$ et al. (1998), there is a
limitation for the observations like ${\rm [Zn/H]+Log}(N_{\rm HI})
\lesssim 21$, which has been explained as dust obscuration (see
also Prochaska \& Wolfe 2002). After careful criticizing the above
dust induced selection bias, Schaye (2001a) proposed a physical
explanation that clouds with $N_{\rm HI}> 10^{22} \cm^{-2}$ will
not appear because in this case, neutral hydrogen will be
converted into molecular hydrogen before reaching such high column
density. However, observations show that in general, the mean
molecular fraction is small in DLAs ( see Ledoux, Petitjean \&
Srianand 2003 and reference therein), and there is no correlation
between the observed amount of H$_2$ and the HI column density.
This implies that the explanation proposed by Schaye (2001a),
although quite attractive, still could not explain many other
observations. In our model, this is because most DLAs with high HI
column densities are absorbers penetrating through disk central
regions where the absorption cross sections are very small. It
should be mentioned that observations show a trend that this upper
limit of HI column density is larger at lower redshifts (SW00),
which, we suggest, could be due to the result of inside-out disk
formation.

If $f(N_{\rm HI},z)$ is known, the number density of DLAs per unit
redshift $N(z) = {\rm d}N/{\rm d}z$ can be calculated through
\begin{equation} \label{eq:dNdz}
N(z) = \frac{{\rm d}X}{{\rm d}z}
       \int_{N_l}^{N_u} f(N_{\rm HI},z){\rm d}N_{\rm HI},
\end{equation}
where $N_l$(= 10$^{20.3} \cm^{-2}$) and $N_u$ are the lower and
upper limits of HI column density distribution for DLAs,
respectively. Meanwhile, the mass density of neutral hydrogen
associated with DLAs can be estimated by
\begin{equation} \label{eq:omegaDLA}
\Omega_{\rm DLA}(z) = \frac{H_0}{c}\frac{\mu m_H}{\rho_{crit}}
       \int_{N_l}^{N_u} N_{\rm HI}f(N_{\rm HI},z){\rm d}N_{\rm
       HI},
\end{equation}
where $m_{H}$ is the mass of the hydrogen atom, $\mu$ is the mean
atomic weight per particle, $\rho_{crit} = 3H_0^2/(8\pi G)$ is the
critical density at present time. Noted that $N(z)$ and
$\Omega_{\rm DLA}$ are dominated by $N_l$ and $N_u$, respectively,
for the power index between $-2$ and $-1$ of $f(N_{\rm HI},z)$.

The model prediction for $N(z=3)$ is about 0.26, in consistence
with the SW00 result according to their fitted formula: $N(z) =
N_0(1+z)^\gamma$, which gives $N(z=3) = 0.256$ for $N_0$ = 0.055
and $\gamma$ = 1.11. It is some how larger than the value given by
Peroux et al.(2003), which is about 0.24. We notice that in Figure
4, model predicted $f(N_{\rm HI})$ is much smaller than
observational ones at low column density. The agreement of number
density $N(z)$ between the model prediction and the observation is
because the model predicted distribution of $f(N_{\rm HI})$ is
flatter, i.e., model predicts more DLA systems with higher column
densities.

Meanwhile, we get the predicted $\Omega_{\rm DLA} \sim 3 \times
10^{-3}$ with the observation result being $\sim 1 \times 10^{-3}$
at redshift 3 (P\'eroux et al. 2003, Boissier et al. 2003). The
reason that the predicted neutral gas density associated with DLAs
is higher than that observed is mainly because the upper limit of
HI column density $N_u$ is larger in our modelled DLAs. Another
possible reason is that we simply assume that all Hydrogen gas is
in atomic form without considering the molecular fraction. The
later one is very complicated because the general transferring
between HI and $\rm H_2$ is still unclear (Ledoux, Petitjean \&
Srianand 2003, Petitjean, Srianand \& Ledoux 2002).

\subsection{SFR density}

Based on the selected DLA sample, we can get the predicted SFR
density contributed by DLAs at $z\sim 3$ and show it as a cross in
Fig. 5 which displays the cosmic SFR density as a function of
redshift resulted from different observations. The most recent
observational results done dy WGP03 for DLAs with $z \gtrsim 2$
based on the $\rm CII^{*}$ absorption lines are also plotted in
the figure as triangles. It can be found that the model prediction
is consistent with observations and supports the ``consensus"
model described by WGP03.

As suggested by WGP03, the SFR densities resulted from DLAs are
similar to that from high redshift luminous galaxies, which are
observed as Lyman Break Galaxies (LBGs). This implies a possible
connection between these two populations as suggested by Schaye
(2001b). On the other hand, Mo, Mao \& White (1999) proposed that
DLAs and LBGs are two distinct populations by the fact that DLAs
are in favor of extended galaxies with large angular momentum
(spin parameter $\lambda$), while LBGs are in favor of compact
systems with small angular momentum. They may exhibit very
different observational properties such as distributions of sizes
and SFRs. So, it is interesting to compare the luminosity
functions of selected DLA host galaxies with that of LBGs.

\begin{figure} [t]
\resizebox{\hsize}{!}{\includegraphics{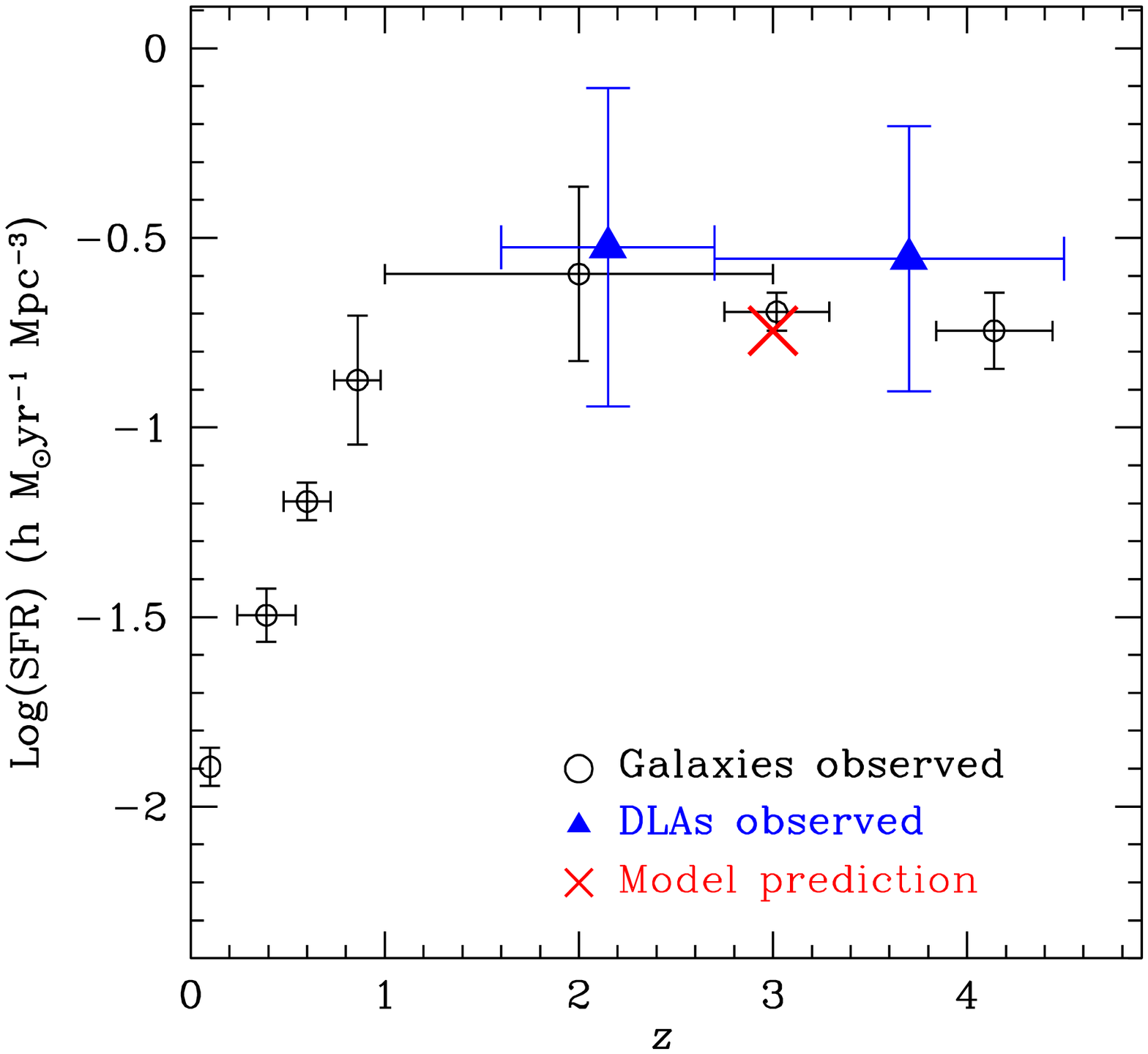}} \caption{Cosmic
SFR density as a function $z$ with the model prediction
contributed by DLAs at $z \sim 3$ as a cross. The open circles and
filled triangles with error bars are the observational results
from galaxy (Barger et al. 2000; Steidel et al. 1999; Lilly et al.
1996) and from DLAs (WGP03), respectively. } \label{Fig5}
\end{figure}

\begin{figure} [t]
\resizebox{\hsize}{!}{\includegraphics{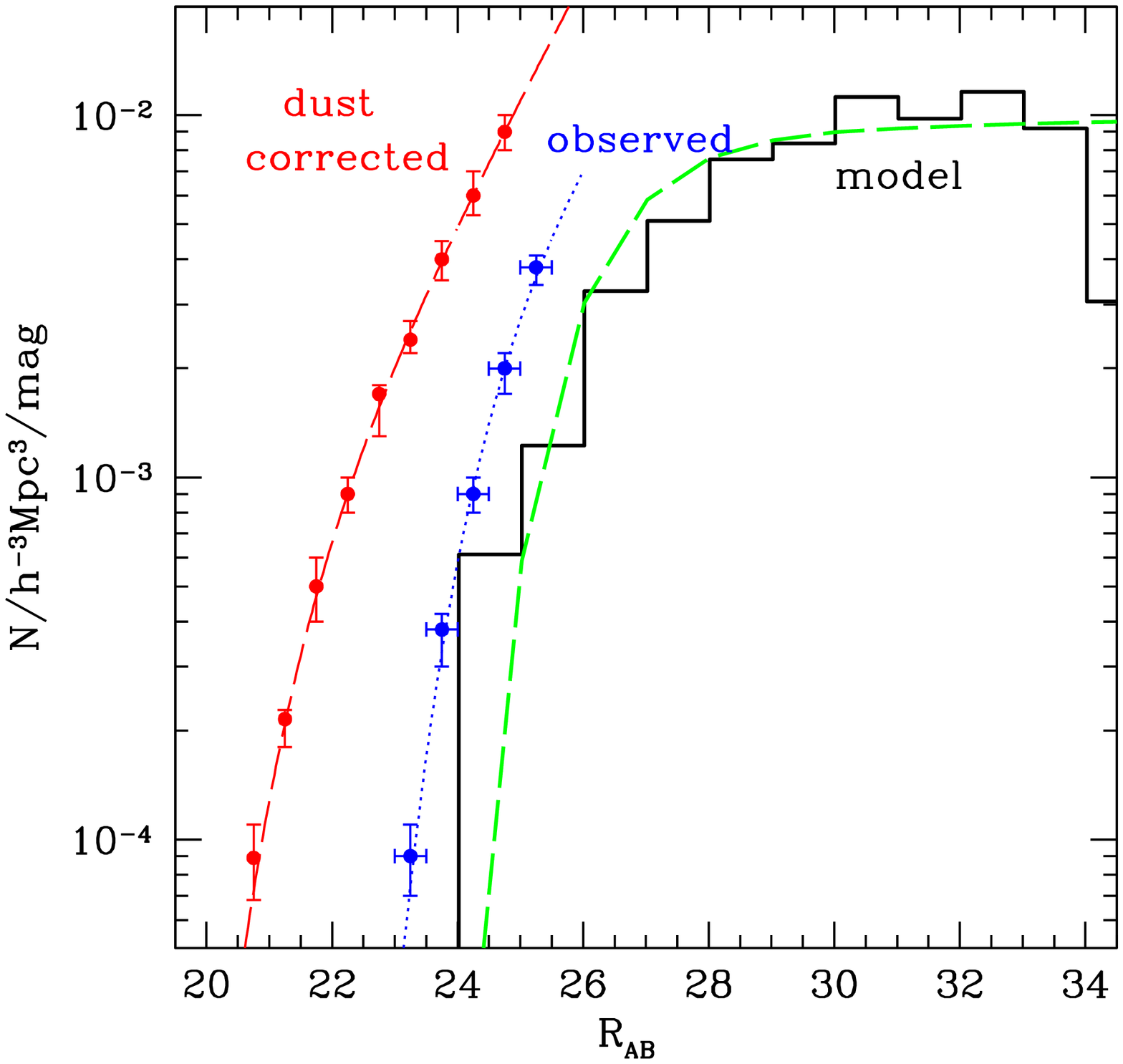}} \caption{UV
luminosity functions with the solid histogram denoting the
modelled DLAs, the dotted and dashed lines denote LBGs without and
with dust-correction which are taken from Aldelberger \& Steidel
(2000). The green line is the least-square fitting of Schechter
function to the modelled histogram, in which the data points are
weighted by a random fluctuation. The fitted parameters are $M_* =
26.1$, $\alpha = -1.01$, $\Phi_* = 1.9\times 10^{-2}$,
respectively. } \label{Fig6}
\end{figure}

In Figure 6, we show, with solid histogram, the predicted UV
luminosity function of selected DLA hosts galaxies. In order to
parameterize the model results, we have fitted a Schechter
function to the modelled histogram, in which the data points are
weighted by a random fluctuation. The fitted parameters are $M_* =
26.1$, $\alpha = -1.01$, $\Phi_* = 1.9\times 10^{-2}$.

The observed and dust-corrected UV luminosity functions of LBGs
are also plotted as full circles with error bars (data from
Aldelberger \& Steidel 2000). Here the prescriptions suggested by
Madau et al.(1998) and Steidel et al.(1999) are adopted to convert
the model SFR distribution of DLA host galaxies to their UV
luminosity function in $\Lambda$CDM cosmology. It can be found
that the typical $R_{\rm AB}$ magnitude of the predicted DLA hosts
is $\sim 30$, which are much fainter than LBGs with typical
$R_{\rm AB}\sim 25$. This implies that a typical DLA host galaxy
in our model has its SFR 100 times smaller than a typical LBG.
Because the number density of DLA host galaxies is about 0.26 (see
previous subsection) which is about 100 times larger than the
observed comoving number density of LBGs, our predictions of
cosmic SFR density contributed by DLAs is similar to that of LBGs
at $z \sim 3$.

Moreover, it can be found from the figure that only a few percent
DLA host galaxies have $R_{\rm AB}$ brighter than 25.5 which is
the current observational threshold for LBG observation at $z \sim
3$. This is consistent with current status about DLA/LBG
connection done by both observationally in $HST$ (M\"{o}ller \&
Warren 1998) and model predictions by Shu (2000). According to the
above discussions, we suggest that DLAs and LBGs should be
physically different populations although chemical evolution
models have shown that DLAs and LBGs are smoothly connected in
their evolutionary history, within which LBGs are galaxies with
shorter star formation timescale like starburst, and DLAs proceed
slower star formation (Shu 2000; Ma \& Shu 2001).

How would dust obscuration (e.g. the Boiss\'e bias) affect the
above results? In order to test this argument, we have applied the
bias proposed in Prantzos \& Boissier (2000) and Hou, Boissier \&
Prantzos (2001), that is to exclude the systems that satisfy $F =
[Zn/H]+log(N_{HI}) > 21$. We found no significant difference for
the luminosity function and also the impact parameter
distribution. There seems to be no evidence, at least in the
present model framework, that a major fraction of the SFR density
could occur in galaxies with significant dust obscuration.

\subsection{Metallicity vs Column Density }

A very unusual property of DLAs observed is that there seems to be
a trend of anti-correlation between their [Zn/H] and HI column
densities, which is independent of redshift as noticed by Boiss\'e
et al.(1998). In fact, as demonstrated by Hou, Boissier \&
Prantzos (2001) and Prochaska \& Wolfe (2002), this
anti-correlation trend exists for almost all the observed
iron-peak elements as well as [Si/H] in DLAs. This could be
possibly due to the existence of dust obscuration in observed DLAs
since the same trends for some elements are also found in the
Milky Way galaxy (Wakker \& Mathis 2000). However, we argue that
this should not be always the case for DLAs, since the trends of
anti-correlation for Zn and Si vs  $N_{\rm HI}$ also exist, while
these two elements are not depleted or only slightly depleted
(Savage \& Sembach 1996). Some other mechanisms should be invoked.

Boiss$\acute{e}$ et al. (1998) claimed that this anti-correlation
is not physical. The absence of DLAs with low metallicities and
low HI column densities could be attributed to the observational
selection effects, i.e., below some level of the HI column density
current spectroscopy is not able to detect metal atoms along
sightlines towards QSOs. And, the lack of DLAs with both high
metallicities and column densities is due to the dust obscuration.
By proper treatments of extinction on their disk models, Prantzos
\& Boissier (2000) have shown that extinction towards background
QSOs increases rapidly as the quantity ${\rm [Zn/H]+ Log}(N_{\rm
HI})$ (which is $\sim$ Zn column density) is greater than 21,
making the background QSOs unobservable in the optically selected
survey. Such kind of interpretation could be tested if, on one
hand, one could be able to detect DLAs towards much fainter QSOs,
and/or on the other hand, more sensitive instruments will become
available so that more metal-poor DLAs with low column densities
could be detected. An alternative is to consider the properties of
DLAs in a complete, radio selected QSO sample, which should not be
influenced by the dust content. Indeed, such a survey has been
done by Ellison et al. (2001). Those authors have surveyed a
sample of DLAs toward radio selected quasars and found no
significant difference in the HI distribution of those DLAs.
Moreover, dust obscuration has also been argued by Prochaska \&
Wolfe (2002), who made a detailed analysis of dust extinction
contained in DLAs. They found that inferred extinction values and
apparent magnitudes imply dust obscuration plays a relatively
minor effect in the DLA analysis at least for $z>2$.

Recent three dimensional SPH simulation of disk galaxy formation
by Churches et al. (2004) have shown that in the modelled disks, a
significant fraction could be optically thick and have higher
column density than observed in DLAs. This implies that background
quasars would be obscured by some disks, producing selection
effect toward the denser absorption systems. But there is no
substantive evidence in HI observations. So, the observed trend of
anti-correlation between metallicity and column density remains as
an open question in DLA studies.

\begin{figure} [t]
\resizebox{\hsize}{!}{\includegraphics{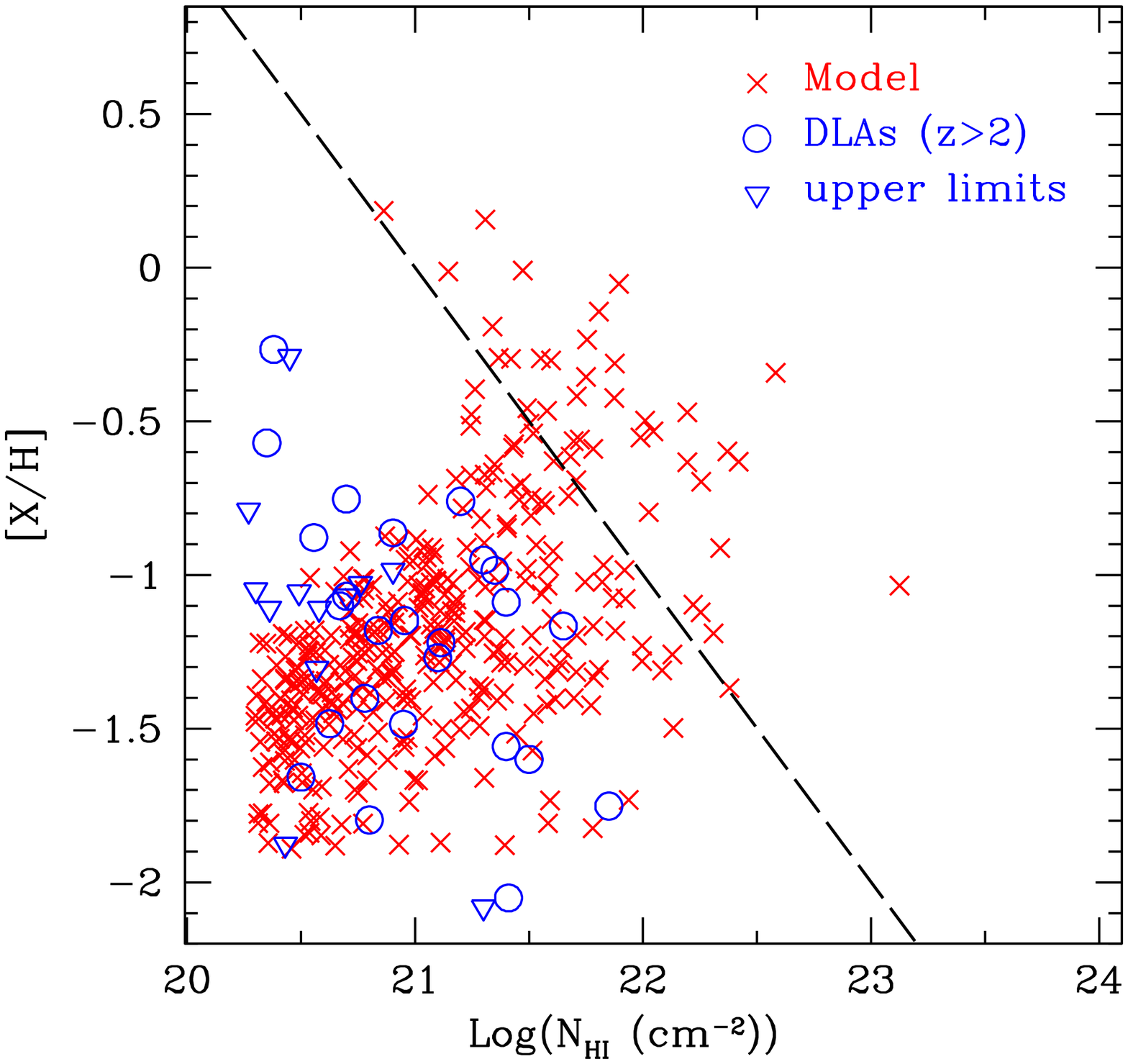}} \caption{The
predicted correlation between metallicity [X/H] and HI column
density $N_{\rm HI}$ for DLAs with crosses and open circles
denoting modelled DLAs and observations for $z>2$, respectively.
The upside-down triangles are those upper/lower limits for $z>2$
DLAs (Kulkarni \& Fall 2002). The long dashed line is ${\rm
[Zn/H]+ Log}(N_{\rm HI}) = 21$. }
\label{Fig7}%
\end{figure}

We examine the predicted correlation between metallicity and HI
column density for the selected DLA sample in Fig.7 with the
observed results of DLAs at $z > 2$ by open circles. As expected,
model predictions show an opposite trend compared with
observations because high gas surface density will have high SFR
which will lead to high metallicity. This is also the main reason
why galaxies always display negative metal gradients. If we apply
the proposed bias of ${\rm [Zn/H]+ Log}(N_{\rm HI}) > 21$ in Fig.7
(long dashed line), to exclude the points above this line, the
difference still exists. This means that the suggestions of
Boiss\'{e} et al (1998) did not help a lot in alleviating the
discrepancy in our model.

Because of the simplification of our model, there still are some
possibilities that could lead to the observed anti-correlation
which are not considered here. First, the absorptions take place
within galactic central regions where HI column densities are low
and metallicities are high. Observations of some spiral galaxies
have shown that in the central regions there exists central HI
depressions, but significant star formations is still ongoing
(Broeils \& van Woerden 1994; Wolfe \& Prochaska 1998; Wong \&
Blitz 2002). It should be noted that the absorption cross
sections, hence the probability, are low for this mechanism.

Another possibility could be the inadequacy of the adopted Schmidt
type star formation prescription. Even for nearby galaxies, the
physical basis of star formation is still poorly known. Observers
have shown various empirical prescriptions for star formation in
spirals (Kennicutt 1998; Rownd \& Young 1999; Wong \& Blitz 2002),
and most of galaxies can be fitted by a Schmidt type law. This has
been widely applied in semi-analytic model of galaxy evolution
(Kauffmann 1996; Boissier \& Prantzos 1999; Chang et al. 1999,
2004). In fact, gas surface density ($\Sigma_g$) includes both the
contributions of HI ($\Sigma_{\rm HI}$) and $\rm H_2$
($\Sigma_{\rm H_2}$) with $\Sigma_{\rm HI}$ being dominant but the
correlation between these two is different from galaxy to galaxy.
Recent observations of star formation regions in nearby galaxies
done by Wong \& Blitz (2002) showed a complex relationship between
SFR and $\Sigma_{\rm HI}$. For their spiral galaxies sample
(biased to molecule-rich galaxies), SFR shows virtually no
correlation with $\Sigma_{\rm HI}$, suggesting a maximum HI column
density around $10^{21} \cm^{-2}$. This is very instructive to the
star formation history for DLAs, where the observed HI column
density seems have an upper limit.

It should be pointed out that we assume in the present paper that
all DLAs are hosted by disks. This seems to be too serious for DLA
population since some observations suggested that DLAs could be
hosted by either disks, spheroids or moving clouds within galactic
halos (Maller et al. 2003). Star formation and chemical enrichment
are quite different for different kinds of hosts, which could lead
to the observed anti-correlation between metallicity and HI column
density for DLAs. For instance, if the few observed (two or three)
points in Fig.7 with very high metallicity ($\rm [X/H] \gtrsim
-1$) and with low HI column density ($N_{\rm HI} \lesssim
10^{21}\cm^{-2}$) are not hosted by disks we assumed, the model
predictions seem to roughly match observations. Furthermore, some
observed points show low metallicity and high HI mass, which could
be gas rich and metal poor galaxies with their star formation
timescales longer than 3Gyr like low surface brightness galaxies.
From this point of view, it seems that the discrepancy between
model predictions and observations could not be very serious. More
observations are needed.

\subsection{Disk mass fraction $m_d$ and star formation timescale}

In the present paper, there is a very important parameter imposed
in our model, which is the disk mass fraction $m_d$. As mention
above, the adoption of eq. (\ref{eq:md}) is based on the
consideration of supernovae feedback which is still poorly
understood. Except $m_d$, the star formation timescale is another
parameter in the model which will influence the final results. In
this subsection, we will discuss the effects of $m_d$ and the star
formation timescale on our obtained results as follows.

Initially, baryon fraction within individual halos is always
considered to be the same as 0.1. At high redshift, significant
fraction of baryons ($\ga 50\%$) can be cooled down radiatively
because of high densities (White \& Frenk 1991). Due to supernovae
feedback, the effective mass fraction which can really form disks
and stars is less. As a test, we assume that the mass fraction of
individual resulting disks is a constant 0.05, i.e., to force
$m_d=0.05$. We find that the resulted $\vcir$ distribution of
modelled DLAs now is dominated by small halos more strongly
because smaller galaxies can host more DLAs due to higher surface
densities without considering the supernovae feedback (see Figure
1(a) ). Other predicted results, such as $\lambda$ distribution,
column density distribution $f(N_{\rm HI})$, $N(z)$, $\Omega_{\rm
DLA}$ and the contributed SFR density, only change slightly.

As a further check, we assume $m_{d0} = 0.05$ in eq. (\ref{eq:md})
with the consideration of supernovae feedback. We find that all
the physical results obtained do not change significantly. It is
mainly because the change of $m_d$ with a factor of 2 will not
change the absorption cross section very much (see eq. (14) of
MMW). The corresponding SFR density will change by a factor less
than 2 (eq. \ref{eq:SFR}), which is about 0.30 in the logarithmic
plot in Figure 5. The uncertainty of $m_d$ should be within the
range we consider above. So, we conclude that the change of $m_d$
within its reasonable range will not change our model results
significantly.

For the effective yield, it will change with the change of
different adoptions of $m_d$ because it is obtained by
best-fitting the modelled distribution of DLA metallicities to
observational ones. Larger is the $m_d$, smaller is the effective
yield sice star formation will be more active for higher gas
surface densities. It can be easy to estimate that the difference
of effective yield for the different adoption of $m_d$ is less
than 20\% according to eqs. (\ref{eq:SFR}) and (\ref{eq:Z}) if the
star formation timescale for DLAs is considered to be some Gyr
(see below).

Moreover, we have also tested the changes of stellar yield for
different adopted ranges of star formation timescales. Following
combinations of timescales, which are within reasonable ranges,
have been tested: $(t_1,t_2) = $(0.5,2.0), (0.5,3.0), (1.0,2.0)
Gyr. It is found that the resulted effective stellar yields were
consistent with each other within the relative error of 15\%. The
general tendency is that the required effective stellar yield
increase when the star formation timescale or the average age of
DLAs decreases. The corresponding resulted $f(N_{\rm HI})$,
$N(z)$, $\Omega_{\rm DLA}$ and SFR densities are also agree with
each other, respectively, within this error.  It means that some
changes of the star formation timescale will not influence our
conclusions.

\section{CONCLUSIONS}

Within the framework of disk galaxy formation theory developed by
Mo, Mao \& White (1998), disks formed in the center of dark matter
halos and are uniquely determined by two parameters: halo circular
velocity $\vcir$ and spin parameter $\lambda$. For a specific
$\Lambda$CDM cosmogony, we can generate a population of galaxies
by Monte Carlo simulations according to the PS formalism of
$\vcir$ and the distribution of $\lambda$ at a given redshift 3.
Star formation and chemical evolution proceed within disks
assuming a typical timescale $1 \sim 3 \rm Gyr$. We select
modelled DLAs according to their observational criterion with the
random inclination being considered, i.e., $N_{\rm HI} \gtrsim
10^{20.3} \cm^{-2}$, to investigate their global properties such
as HI column density and metallicity. The main results are
summarized as follows:

We get the effective yield $y = 0.25Z_{\odot}$ by best-fitting
resulted metallicity distributions to the observed one of DLAs.
Our model can well reproduce the observed metallicity distribution
of DLAs. The relatively low value of the effective yield indicates
that galactic winds could play an important role during star
formation in disks which could relate to the kinematics of DLAs
observed.

On the basis of simulated DLAs, we have calculated the frequency
distribution of HI column densities which is a bit flatter than
the observed ones (SW00). This could be due to the model
limitations. For instance, disk instability for star formation
criterion is not considered (Ma \& Shu 2001). The number density
and mass density contributed by DLAs at $z \sim 3$ are also
discussed, in which the predicted number density agrees well with
observations and the predicted mass density contribution is
larger. Our model suggests that DLAs could naturally arise in a
$\Lambda$CDM universe from radiatively cooled gas in dark matter
halos.

Furthermore, the predicted SFR density at $z \sim 3$ contributed
by DLAs is consistent with the most recent observations (WGP03).
Because the SFR density contributed by DLAs has the same order as
that by LBGs, it is interesting to compare the UV luminosity
functions of these two populations. We find that the typical DLA
host galaxy is much fainter than LBGs. It implies that very few
DLA hosts can be observed as LBGs which is consistent with current
observations. We suggest that these two population of galaxies
should be physically different.

It should be pointed out that model predicted correlation between
metallicity and HI column density for DLAs cannot match
observations well, even if the proposed observational bias
suggested by Boiss\'e et al. (1998) is taken into account. We
suggest that the observed trend of anti-correlation could most
probably be physical. Some possible mechanisms with the model
simplicities were discussed. Still, more observations are needed
to clarify this trend in the future.

\section*{Acknowledgements}

We are grateful to B. M\'enard, W.P. Lin, H.J. Mo, and D.H. Zhao
for useful discussions on the subject. S. Boissier is specially
thanked for very useful comments and helping us with the data.
Many thanks should be given to the referee whose critical
arguments greatly improve the manuscripts. This work is supported
by NSFC10173017, 10133020, 10073016, NKBRSF 1999075404 and
NSC91-2112-M008-036. CS acknowledges the financial support of NSC
for a visit to NCU and the kind hospitality during the stay in
NCU.

\def\cjaa{CJAA}

{}

\end{document}